\documentclass[11pt,sort&compress]{elsarticle}
\usepackage[paper=a4paper,top=24mm, bottom=26mm, left=26mm, right=26mm]{geometry}
\makeatletter
\def\ps@pprintTitle{%
 \let\@oddhead\@empty
 \let\@evenhead\@empty
 \def\@oddfoot{\centerline{\thepage}}%
 \let\@evenfoot\@oddfoot}
\makeatother
\usepackage{natbib}
\usepackage[hyperref]{xcolor}
\definecolor{darkgreen}{rgb}{0.01, 0.75, 0.24}
\definecolor{darkblue}{HTML}{2B66D3}
\usepackage[colorlinks=true,
            linkcolor=darkblue,
            urlcolor=darkblue,
            citecolor=darkblue,linkcolor=darkblue,hyperfootnotes=true]{hyperref}

\usepackage{amsmath}
\usepackage{amsfonts}
\usepackage{slashed}
\usepackage{accents}
\usepackage{amssymb}
\usepackage{mathrsfs}
\usepackage{mathtools}
\usepackage[utf8]{inputenc}
\usepackage[T1]{fontenc}
\let\oldbibliography\thebibliography
\renewcommand{\thebibliography}[1]{%
  \oldbibliography{#1}%
  \setlength{\itemsep}{1.4pt}%
}

\usepackage[cal=boondox,calscaled=1]{mathalfa}
\DeclareMathAlphabet{\bbvar}{U}{BOONDOX-ds}{m}{n}

\usepackage{psfrag}
\usepackage{pstool}
\usepackage{caption}

\usepackage{tabularx}
\usepackage{makecell}
\usepackage{multirow}
\usepackage{pbox}
\usepackage{graphicx}
\usepackage{subfigure}
\newcommand{\di}{\mathrm{d}}
\usepackage{tensor}
\newcommand{\ou}[3]{\tensor{#1}{^{#2}_{#3}}}

\newcommand{\I}{\mathrm{i}} 
\newcommand{\E}{\mathrm{e}} 
\newcommand{\C}{\mathbb{C}}

\newcommand{\R}{\mathbb{R}}

\newcommand{\eref}[1]{(\ref{#1})}

\usepackage{siunitx}

\DeclareMathAlphabet{\bbgreek}{U}{bbold}{m}{n}

\newcommand{\mtext}[1]{\text{\it #1}}
\usepackage{tikz-cd}

\newcommand{\sh}{\operatorname{sh}}
\newcommand{\ch}{\operatorname{ch}}
\renewcommand{\th}{\operatorname{th}}
\newcommand\vpm{\mathbin{\vcenter{\hbox{
  \oalign{\hfil$\scriptstyle+$\hfil\cr
          \noalign{\kern-.3ex}
          $\scriptscriptstyle({-})$\cr}}}}}
\DeclareMathAlphabet{\sfit}{OT1}{fos}{sb}{it}
\DeclareMathAlphabet{\mathsf}{OT1}{fos}{sb}{n}

\definecolor{darkgreen}{rgb}{0.01, 0.75, 0.24}
\definecolor{darkblue}{HTML}{2B66D3}

\usepackage[multiple, flushmargin]{footmisc}
\let\originalleft\left
\let\originalright\right
\renewcommand{\left}{\mathopen{}\mathclose\bgroup\originalleft}
\renewcommand{\right}{\aftergroup\egroup\originalright}

\newcommand{\dbarvar}{{\mathrm{d}\mkern-7.5mu\lower.18ex\hbox{$\textasciitilde$}\mkern-1.5mu}}

\renewcommand{\emph}[1]{{\it #1}}

\begin{document}

\begin{abstract}
\noindent 
Recently, we introduced a non-perturbative quantization of impulsive gravitational null initial data. In this note, we investigate an immediate physical implication of the model. One of the quantum numbers is the total luminosity carried to infinity. We show that a transition happens when the luminosity reaches the Planck power $\mathcal{L}_{\mathrm{P}}$. Below $\mathcal{L}_{\mathrm{P}}$, the spectrum of the radiated power is discrete. Above the Planck power, the spectrum is continuous. A physical state that lies in the continuous spectrum consists of a superposition of kinematical states in which the shear is unbounded from above. We argue that such states are unphysical because they contain caustics that are in conflict with the  falloff conditions at asymptotic infinity.
\end{abstract}
\title{Evidence for Planck Luminosity Bound in Quantum Gravity}
\author{Wolfgang Wieland\footnote{wolfgang.wieland@fau.de}}
\address{Institute for Quantum Gravity, Theoretical Physics III, Department of Physics\\Friedrich-Alexander-Universität Erlangen-Nürnberg, Staudtstra\ss e 7, 91052 Erlangen, Germany\\{\vspace{0.5em}\normalfont November 2024}
}

\maketitle

\section{Introduction}

\noindent It has been often argued that the Planck power
\begin{equation}
\mathcal{L}_{\mathrm{P}}=\frac{m_{\mathrm{P}}c^2}{t_{\mathrm{P}}}=c^5/G\approx 3,63\times\SI{e52}{\watt}\label{PlanckL}
\end{equation}
places an upper bound on the gravitational wave luminosity, see e.g.\ \cite{Misner:1973prb,Cardoso:2018nkg,Jowsey:2021gny}. The perhaps simplest argument in favour of this idea can be found in Misner, Thorne, Wheeler \cite{Misner:1973prb}. Consider the virial theorem, i.e.\ $E_{\mathrm{kin}}\sim M\omega^2R^2 \sim GM^2/R$, and the quadrupole formula for the gravitational wave luminosity, i.e.\ $\mathcal{L}_{\mathrm{GW}}\sim \mathcal{L}_{\mathrm{P}}^{-1}(MR^2\omega^3)^2$, where $M$ is the mass of the system, $R$ denotes its spatial extension and $\omega$ is the frequency at which it oscillates. Since the emission can only happen when  $R$ is greater than the Schwarzschild radius $r_S=2GM/c^2$, we obtain $\mathcal{L}_{\mathrm{GW}}\lesssim\mathcal{L}_{\mathrm{P}}$. \hyperref[tab1]{Table 1} lists a few astrophysical sources. So far, no observation exceeds the Planck luminosity, but there is also no proof that a sufficiently large class of initial data sets for e.g.\ asymptotically flat solutions of Einstein's equations would satisfy such a bound on the gravitational luminosity.\smallskip 

It is also unlikely that there is such a bound in higher dimensions. In $D>4$ spacetime dimensions, the Planck power depends not only on $G$ and $c$, but also on $\hbar$, which can not appear at the classical level. It is no surprise therefore that it is in $D=4$ alone that we can have an equation for the peak luminosity of two coalescing black holes, which 
depends only on dimensionless observables, such as the mass ratio $\eta=m_1m_2/(m_1+m_2)^2$ and dimensionless spin components \cite{Keitel:2016krm,PhysRevLett.120.111101}. In higher dimensions, we need an additional length scale \cite{Cardoso:2002pa}, e.g.\ the frequency at which the system oscillates. This makes it unrealistic that there is a bound on the gravitational wave luminosity when $D>4$.\smallskip

Recently, we introduced a non-perturbative quantization of impulsive gravitational null initial data in $D=4$, see  \cite{Wieland:2024dop}.  The proposal relies on the geometry of light-like boundaries, which simplifies the Hamiltonian analysis \cite{Reisenberger:2012zq,Reisenberger:2018xkn,Ciambelli:2023mir,Wieland:2021vef}. In this note, we investigate the role of the Planck power $\mathcal{L}_{\mathrm{P}}$ in the model. The analysis is based on a combination of non-perturbative and semi-classical techniques. The key result is that $\mathcal{L}_{\mathrm{P}}$ separates the spectrum of the emitted power into a discrete and continuous part. This happens upon adding a parity-violating $\gamma$-term \cite{Samuel:1987td,Jacobson:1987yw,holst} to the gravitational action in the bulk. This term creates an effective compactification of certain directions on phase space, which in turn affects the spectra of physical observables \cite{Wieland:2017cmf,Wieland:2024dop,Freidel:2020ayo,Freidel:2005sn}. The Holst action \cite{holst}, which underpins this model, is given by 
\begin{align}
S[A,e]=\frac{1}{16\pi G}\int_{\mathcal{M}_4}&\left[\tfrac{1}2\ou{\epsilon}{\alpha\beta}{\alpha'\beta'}-\tfrac{1}{\gamma}\delta^{[\alpha}_{\alpha'}\delta^{\beta]}_{\beta'}\right]e_\alpha\wedge e_\beta\wedge F^{\alpha'\beta'},\label{gammaactn}
\end{align}
where $F=\di A+\tfrac{1}{2}[A,A]$ is the curvature of the $\mathfrak{so}(1,3)$ connection, $e^\alpha$ is the co-tetrad and $\gamma$ is the Barbero--Immirzi parameter \cite{Barbero1994,Immirziparam}. Since $\mathcal{L}_{\mathrm{P}}\rightarrow\infty$ as $G\rightarrow 0$, we also note that it seems implausible to find a bound on the gravitational wave luminosity from a perturbative quantisation, where we have a formal perturbative expansion with respect to $\varkappa=\sqrt{8\pi G/c^3}$. \smallskip

In $D=4$, Planck's constant cancels from $\mathcal{L}_{\mathrm{P}}=m_{\mathrm{P}}c^2/t_{\mathrm{P}}$. If quantum theory is responsible for a bound on gravitational wave luminosity, this can only be the result of a subtle effect in which what matters is not the actual numerical value of $\hbar$, but rather that $\hbar\neq 0$. This is reminiscent of statistical mechanics. To compute the microcanonical entropy, we divide the classical phase space of $N$ indistinguishable particles into bins of size $(2\pi\hbar)^{3N}$ and count the number of mircostates compatible with the macroscopic observables $(U,V)$. Smaller bins would be in violation of the Heisenberg uncertainty relations. The resulting microcanonical entropy $S(U,V,N)$ is finite. As we send $\hbar\rightarrow 0$, the entropy diverges, but this logarithmic divergence is invisible when considering processes at fixed particle number. A different effect, but similar in logic, appears in string theory. The appearance of the Weyl anomaly is an $\hbar\neq 0$ effect. The anomaly vanishes if we add extra dimensions. The string theorists' prediction that we live in a world, in which there are seven extra dimensions hidden from us, is a quantum effect clearly independent of the actual numerical value of Planck's constant.\smallskip
\smallskip

In addition, there is an important lesson from gravity in $D=3$ spacetime dimensions. In $D=3$, it is the Planck mass $m_{\mathrm{P}}=c^2/G_{3}$ rather than the Planck luminosity that is independent of $\hbar$. Whereas there is no proof in $D=4$ for a bound on the luminosity of a gravitational wave,\footnote{Ref. \cite{Jowsey:2021gny} contains a counter example that demonstrates that there is no such bound for spin-0 matter in classical relativity. The authors consider a spherically symmetric star that radiates null dust into the environment. By carefully tuning the parameters of the evaporating star, it is possible to make the luminosity arbitrarily large. However, this is not in contraction with a conjectured bound on the \emph{gravitational wave} luminosity. A spherically symmetric star does not radiate gravitational waves.}  we do have a simple proof in $D=3$ that the asymptotic mass $m_{\mathrm{ADM}}$ is bounded from both sides \cite{Deser1984220,Ashtekar:1993ds}, i.e.\ $0\leq m_{\mathrm{ADM}}<m_{\mathrm{P}}/4$. 
At the quantum level, the resulting Hilbert space is constructed from a $q$-deformation of the local De\,Sitter algebra \cite{Witten:1988hc,Meusburger2008,HamquantCS}.  If we then couple point particles to the topological state sum models for quantum gravity in $D=3$ and take a limit in which we send $\hbar\rightarrow 0$ but keep the ratio between the Planck length and the Hubble radius fixed, we end up with a constantly curved momentum space \cite{Girelli:2004md,Amelino-Camelia:2012zkr,Amelino-Camelia:2003ezw,Freidel:2003sp}. This is a very subtle quantum effect in which $\hbar$ alters the \emph{classical} algebra of observables. We will argue below that a similar mechanism exists in 3+1 dimensions, with the Planck power playing the role that the Planck mass has in 2+1 dimensions. Roughly speaking, luminosity is the ratio of two observables, see \eref{critshear1} below. In the ratio, $\hbar$ cancels and we obtain a non-trivial effect as $\hbar\rightarrow 0$.%

\begin{table}{\small
\begin{tabularx}{35.5em}[c]{p{8em} p{14.6em} r r}\Xhline{1pt}
{\it source} & {\it note} &{\it approx.\ power\,}&\phantom{\Large{I}}{\it ref.}\\[0.1em]\Xhline{1pt}
{\it Sun} &  {\it G-type main sequence star} & $\SI{e26}{\watt}$&\phantom{\Large I}\cite{SolSystemUnits}\\[0.1em]
{\it $\eta$ Carinae} & {\it Luminous blue variable star} & $\SI{e33}{\watt}$&\phantom{\Large I}\cite{EtaCarinae}\\[0.1em]\hline
{\it FRBs} &  {\it Fast radio bursts} & $10^{35}$---$\SI{e38}{\watt}$&\phantom{\Large I}\cite{FRB_power}\\[0.1em]
{\it QSO J0529--4351} &  {\it Quasar} & $\SI{e41}{\watt}$&\phantom{\Large I}\cite{Wolf:2024aa}\\[0.1em]
{\it GRB 221009A} &  {\it Brightest gamma ray burst to date} & $\SI{e47}{\watt}$&\phantom{\Large I}\cite{Frederiks_2023}\\[0.1em]
{\it GW 170817} &  {\it Binary neutron star merger} & $\gtrsim\SI{e48}{\watt}$&\phantom{\Large I}\cite{PhysRevX.9.031040}\\[0.1em]
{\it GW 170729} &  {\it Binary black hole merger} & $\SI{e49}{\watt}$&\phantom{\Large I}\cite{PhysRevX.9.031040}\\[0.1em]\hline
{\it Planck scale} &  {\it $c^5/G$} & $\SI{e52}{\watt}$&\phantom{\Large I}\\[0.1em]\Xhline{1pt}
\end{tabularx}\vspace{0.2em}
}
\centering\caption{Comparison between the Planck power and the luminosity of known astrophysical sources.}\label{tab1} 
\end{table}

\section{Phase space of impulsive data}
\noindent In the following, we consider the phase space of a pulse of gravitational null initial data on a null hypersurface $\mathcal{N}_3\simeq[-1,1]\times S_2$ in the boundary of a compact space-time region $\mathcal{M}_4$, i.e.\ $\mathcal{N}_3\subset\partial\mathcal{M}_4$. Notice that $\mathcal{N}_3$ has itself a boundary $\partial\mathcal{N}_3=\mathcal{C}_+\cup\mathcal{C}_-$, consisting of the initial and final cuts $\mathcal{C}_\pm\simeq S_2$.  Since $\mathcal{N}_3$ is a null hypersurface, we can introduce now a co-dyad $(\ou{e}{1}{a},\ou{e}{2}{a})\in\Omega^1(\mathcal{N}_3:\R^2)$, intrinsic to $\mathcal{N}_3$, that diagonalizes the signature $(0$$+$$+)$ metric 
\begin{equation}
q_{ab}:=\varphi^\ast_{\mathcal{N}_3} g_{ab}=\delta_{ij}\ou{e}{i}{a}\ou{e}{j}{b},
\end{equation}
where $\delta_{ij}$ is the Kronecker delta and $\varphi^\ast_{\mathcal{N}_3} g_{ab}$ is the pull-back of the spacetime metric $g_{ab}$ in $\mathcal{M}_4$ to the boundary $\mathcal{N}_3$. 
Any such co-dyad can be parametrized by a conformal factor $\Omega$ and an $SL(2,\R)$ holonomy $S$,
\begin{equation}
\ou{e}{i}{a}=\Omega\,\ou{S}{i}{m}{}^{\circ}\ou{e}{m}{a},\label{e-def}
\end{equation}
where ${}^{\circ}\ou{e}{m}{a}$ is a fiducial dyad, which we keep fixed once and for all. A possible choice is 
${}^\circ\ou{e}{1}{a}=\partial_a\vartheta$, ${}^\circ\ou{e}{2}{a}=\sin(\vartheta)\partial_a\varphi$, where $(\vartheta,\varphi)$ are standard spherical coordinates that are Lie dragged along the null generators. \hyperref[fig2]{Figure 1} illustrates the resulting parametrisation. \smallskip

\begin{figure}[h]
\centering
\subfigure[conformal factor]{
       \centering
       \includegraphics[scale=0.4]{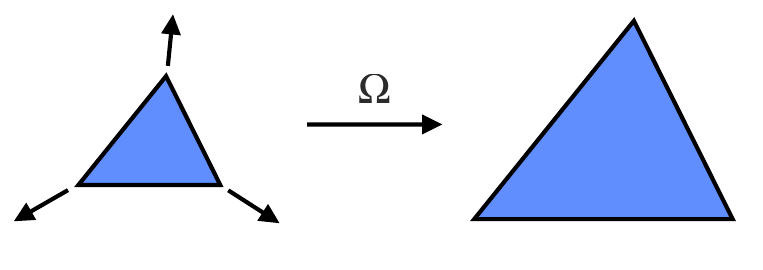}
       \label{fif:}
   }
    \subfigure[shape modes and $U(1)$ gauge invariance]{
        \centering
        \includegraphics[scale=0.4]{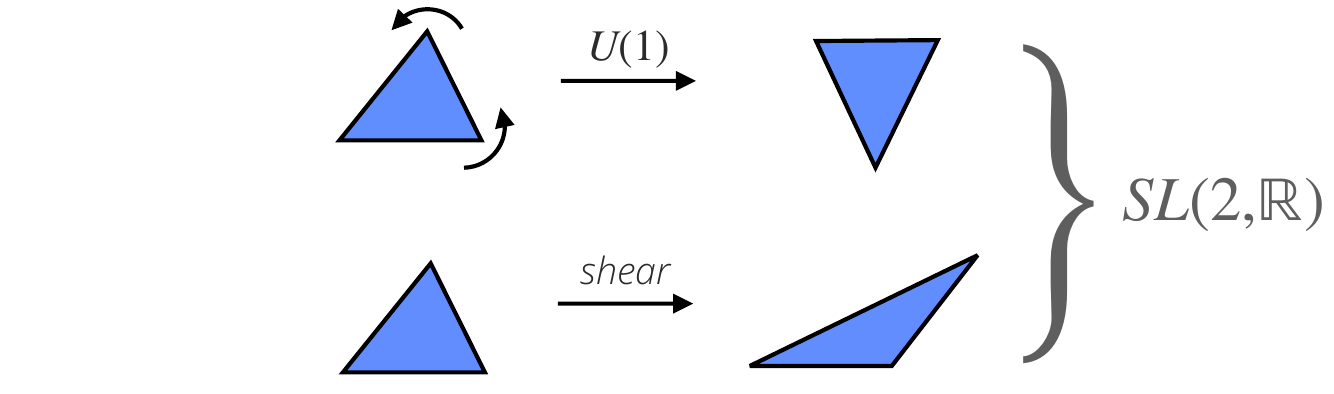}
        \label{fig:ADM2}
    }
    \caption{A null boundary carries a degenerate signature $(0$$+$$+)$ metric, which is obtained by taking the pull-back of the four-dimensional line element to the boundary. The null geometry is effectively two-dimensional. The two-dimensional geometry can be parametrised by an overall conformal factor (left panel) and the residual shape modes taking values in $SL(2,\R)/U(1)$ (right panel).    }
    \label{fig1}
\end{figure}

Next, we introduce a time coordinate $\mathcal{U}:\mathcal{N}_3\rightarrow[-1,1]$ and extend, thereby,  the angular coordinates $(\vartheta,\varphi)$ into a three-dimensional coordinate system of $\mathcal{N}_3$. The resulting vector field $\partial_{\mathcal{U}}$ is null. We assume it to be future pointing. The boundary conditions are
\begin{equation}
\text{for}\quad\partial\mathcal{N}_3=\mathcal{C}_+\cup\mathcal{C}_-^{-1}:\mathcal{U}\big|_{\mathcal{C}_\pm}=\pm 1.\label{Ubndry}
\end{equation}
 In the absence of additional structure, such as a preferred foliation, symmetries or asymptotic boundary conditions, there is no natural torsionless and metric compatible covariant derivative in $T\mathcal{N}_3$, see e.g.\ \cite{Ashtekar:2024mme}.  A natural derivative $D_a$ exists on the extended vector bundle $\bigcup_{p\in\mathcal{N}_3}\{p\}\times T_{p}\mathcal{M}_4$, where it is induced from the bulk, i.e.\ $D_a=\varphi^\ast_{\mathcal{N}_3}\nabla_a$, given  the usual Levi-Civita covariant derivative $\nabla_a$ in $(\mathcal{M}_4,g_{ab})$. The corresponding connection, which depends on both the extrinsic and intrinsic geometry of $\mathcal{N}_3$, is the metric analogue of the self-dual Ashtekar connection \cite{newvariables}. \smallskip

There are infinitely many clock variables $\mathcal{U}$ that satisfy the boundary condition \eref{Ubndry}. To pick a unique representative, we impose the gauge condition
\begin{equation}
\partial^a_{\mathcal{U}}D_a\partial_{\mathcal{U}}=-\Omega^{-1}\frac{\di}{\di\mathcal{U}}[\Omega]\partial_{\mathcal{U}}.
\end{equation}
Upon choosing this gauge, the residual constraints simplify \cite{Wieland:2021vef}. We are left with a transport equation for the $SL(2,\R)$ holonomy and the Raychaudhuri equation \cite{PhysRev.98.1123} for $\Omega^2$. The Raychaudhuri equation becomes
\begin{align}
\frac{\di^2}{\di\mathcal{U}^2}\Omega^2&=-2\sigma\bar{\sigma}\Omega^2,\label{ham-cons}
\end{align}  
where $\sigma$ is the shear of the null generators $\partial_{\mathcal{U}}$ of $\mathcal{N}_3$. The shear $\sigma$ is charged under a $U(1)$  symmetry sending the $SL(2,\R)$ holonomy $S$ introduced in \eref{e-def} into $\tilde{S}=\E^{J\psi}{S}$, where $\psi$ is a $U(1)$ angle. Under this symmetry, $\sigma$ is mapped into $\tilde{\sigma}=\E^{2\I\psi}\sigma$, thus $\sigma$ is a spin-2 field intrinsic to the null boundary.  In here, the $2\times 2$ matrix $J$ is an $U(1)$ generator that preserves the internal metric $\delta_{ij}$, i.e.\ $\delta_{ik}\ou{J}{k}{j}+\delta_{kj}\ou{J}{k}{i}=0$ and satisfies $J^2=-\bbvar{1}$. To fix the $U(1)$ gauge redundancy, we introduce the decomposition
\begin{equation}
S=\E^{\Delta J}HS_-,
\end{equation}
where $\Delta:\mathcal{N}_3\rightarrow[0,2\pi)\,\mathrm{mod}\,2\pi$ is an $U(1)$ angle with boundary condition $\Delta(\mathcal{U}=-1,\vartheta,\varphi)=0$. In addition, $S(\mathcal{U}=-1,\vartheta,\varphi)=S_-(\vartheta,\varphi)$ is the initial condition at $\mathcal{C}_-$. The $SL(2,\R)$ element $H$, on the other hand, depends only on the profile of the shear. It  is the solution of a transport equation with fixed initial conditions, i.e.\
\begin{align}
\frac{\di}{\di\mathcal{U}}H&=\left(\sigma\bar{X}+\bar{\sigma}X\right)H,\qquad H(\mathcal{U}=-1,\vartheta,\varphi)=\bbvar{1}.\label{hol-eq}
\end{align}
In here, we introduced a decomposition of $\mathfrak{sl}(2,\R)$ into translational components $X$ and $\bar{X}$ and the $U(1)$ generator $J$ with $\mathfrak{sl}(2,\mathbb{R})$ commutation relations
\begin{equation}
[J,X]=-2\I X,\;[J,\bar{X}]=+2\I \bar{X},\; [X,\bar{X}]=\I J.\label{Lie-basis}
\end{equation}
In the interior of $\mathcal{N}_3$, the value of $\Delta$ can be gauged to zero. At the upper boundary $\mathcal{C}_+$
it can not. The boundary data $S_-(\vartheta,\varphi)$ and $\Delta_+(\vartheta,\varphi)=\Delta(\mathcal{U}=1,\vartheta,\varphi)$ are an example of gravitational edge modes \cite{Balachandran:1994up,PhysRevD.51.632, Barnich:2011mi, Donnelly:2016auv, Carrozza:2021gju, Freidel:2021cjp,Freidel:2020xyx, Freidel:2020svx,Freidel:2023bnj, Wieland:2020gno,Wieland:2016exy}.\smallskip

The shear $\sigma:\mathcal{N}_3\rightarrow \C$ is unconstrained. It determines the free radiative data along $\mathcal{N}_3$. To describe the quantum geometry of a single pulse of radiation, we consider only those configurations on phase space, where $\sigma$ is constant along the null generators of $\mathcal{N}_3$. We can then integrate the Raychaudhuri equation and obtain
\begin{align}\nonumber
\Omega^2(\mathcal{U},\vartheta,\varphi)=\frac{E_+(\vartheta,\varphi)+E_-(\vartheta,\varphi)}{2}&\frac{\cos\left(\sqrt{2\sigma\bar{\sigma}}\,\mathcal{U}\right)}{\cos\left(\sqrt{2\sigma\bar{\sigma}}\right)}+\\
+\frac{E_+(\vartheta,\varphi)-E_-(\vartheta,\varphi)}{2}&\frac{\sin\left(\sqrt{2\sigma\bar{\sigma}}\,\mathcal{U}\right)}{\sin\left(\sqrt{2\sigma\bar{\sigma}}\right)},\label{Omprofile}
\end{align}
where $E_\pm(\vartheta,\varphi)=\Omega^2(\mathcal{U}=\pm1,\vartheta,\varphi)$ is the free corner data of the area element at the initial and final cross section of $\mathcal{N}_3$. In summary, there is the standard free radiative data along $\mathcal{N}_3$ given by the shear $\sigma$ and in addition, there is the corner data at $\mathcal{C}_\pm$. The corner data consists of the area element $E_\pm$ at the two cuts, the initial data $S_-$ for the transport equation for the $SL(2,\R)$ holonomy and a global $U(1)$ angle $\Delta_+$ that measures the $U(1)$ parallel transport along the null generators from $\mathcal{C}_-$ to $\mathcal{C}_+$. Further details can be found in  \cite{Wieland:2024dop}.\smallskip

The symplectic potential for the Holst action  \eref{gammaactn} on a null boundary has been computed in \cite{Wieland:2021vef,Wieland:2024dop}.
 The contribution from the Barbero--Immirzi parameter can be absorbed back into a boundary term. In terms of the $SL(2,\R)$ variables, we obtain
 \begin{align}
\Theta_{\mathcal{N}}&=-\frac{1}{16\pi{\gamma} G}\int_{\partial{\mathcal{N}}}d^2v_o\,\Omega^2\,\operatorname{Tr}\big(J\bbvar{d}SS^{-1}\big)+\nonumber\\
&\quad-\frac{1}{8\pi G}\int_{\mathcal{N}}\di\mathcal{U}\wedge d^2v_o\,\Omega^2\,\operatorname{Tr}\big((\sigma\bar{X}+\bar{\sigma}X)\bbvar{D}(HS_-)(HS_-)^{-1}\big)\nonumber+\\
&\quad-\frac{1}{ 8\pi G}\int_{\mathcal{N}}\di\mathcal{U}\wedge d^2v_o\,\bbvar{d}\mathcal{U}\Big(\frac{\di^2}{\di\mathcal{U}^2}\Omega^2+2\sigma\bar{\sigma}\Omega^2\Big),\label{symplpot}
\end{align}
where $d^2v_o={}^{\circ}e^1\wedge {}^{\circ}e^2$ is the fiducial area element and $\bbvar{D}=\bbvar{d}-\bbvar{d}\mathcal{U}\frac{\di}{\di\mathcal{U}}$ is a dressed differential on phase space, see \cite{Gomes:2016mwl,Gomes:2018shn,Carrozza:2021sbk,PhysRevD.108.106022,Wieland:2024dop}. The symplectic potential \eref{symplpot} determines the Poisson brackets for null initial for the $\gamma$-action \eref{gammaactn}. The impulsive null initial data that we introduced above is parametrized by the area densities at the two cuts $E_\pm$, the $\C$-valued shear $\sigma$, the $SL(2,\R)$ element $S_-$ and the $U(1)$ angle $\Delta_+$. All these variables are constant along $\mathcal{U}$, but may have a non-trivial angular dependence. The resulting parameter space for all $(E_\pm,\sigma,S_-,\Delta_+)$ determines a submanifold of phase space. By taking the pull-back on phase space of the presymplectic potential \eref{symplpot} to this sumbmanifold, we obtain canonical commutation relations for any such impulsive data \cite{Wieland:2024dop}. The canonical variables contain two sets of harmonic oscillators $a$, $b$. In addition, there is an $SL(2,\R)$ group-valued configuration variable $U$, which is related to $S_-$, and its conjugate momentum, which splits into $\mathfrak{sl}(2,\R)$ components $L$ and $c, \bar{c}$ with respect to the $(J,X,\bar{X})$ basis introduced in \eref{Lie-basis}. 
The relation between the canonical variables and the impulsive boundary data is determined by two sets of equations, see \cite{Wieland:2024dop} for a detailed calculation. First of all, we have
\begin{align}
U&=\E^{\gamma\ln\left(\tan\left(\sqrt{2\sigma\bar{\sigma}}\right)/\sqrt{2\sigma\bar\sigma}\right) J}S_-,\\
a&=\frac{\sqrt{E_+}}{\sqrt{8\pi\gamma G}}\ch\bigl(2\sqrt{\sigma\bar{\sigma}}\bigr)\E^{-\I\left[\Delta_++2\gamma\ln\left(\cos\left(\sqrt{2\sigma\bar{\sigma}}\right)\right)\right]},\\
b&=\frac{\sqrt{E_+}}{\sqrt{8\pi\gamma G}}\sh\bigl(2\sqrt{\sigma\bar{\sigma}}\bigr)\E^{\I\left[\Delta_++\phi+2\gamma\ln\left(\frac{\sin\left(\sqrt{2\sigma\bar{\sigma}}\right)}{\sqrt{2\sigma\bar{\sigma}}}\right)\right]},
\end{align}
where $\Delta_+(\vartheta,\varphi)=\Delta(\mathcal{U}=1,\vartheta,\varphi)$ and $\phi:\sigma=|\sigma|\E^{\I\phi}$ are $U(1)$ angles. The $SL(2,\R)$ element $U$ determines additional corner data that parametrize the shape degrees of freedom of the signature $(0$$+$$+)$ metric at the boundary. The overall scale of the boundary metric is set by the $U(1)$ generator $L$ and the norm of the oscillator variables $a$ and $b$. We obtain
\begin{align}
E_-+E_+&=16\pi\gamma G\left(L+a\bar{a}\right),\\
E_--E_+&=16\pi\gamma G\left(L+b\bar{b}\right).
\end{align}
The quotient of the two oscillators determines the shear 
\begin{equation}
\th\bigl(2\sqrt{\sigma\bar{\sigma}}\bigr)=\sqrt{\frac{\bar{b}b}{\bar{a}a}}.\label{shear-def}
\end{equation}
The only non-vanishing brackets among the fundamental variables are
\begin{subequations}\begin{align}
\big\{a(\boldsymbol{z}),\bar{a}(\boldsymbol{z}')\big\}&=\I\,\delta^{(2)}(\boldsymbol{z}|\boldsymbol{z}'),\\
\big\{b(\boldsymbol{z}),\bar{b}(\boldsymbol{z}')\big\}&=\I\,\delta^{(2)}(\boldsymbol{z}|\boldsymbol{z}'),\\
\big\{c(\boldsymbol{z}),\bar{c}(\boldsymbol{z}')\big\}&=2\,\I\,\delta^{(2)}(\boldsymbol{z}|\boldsymbol{z}')\,L(\boldsymbol{z}),\\
\big\{L(\boldsymbol{z}),{c}(\boldsymbol{z}')\big\}&=-\I\,\delta^{(2)}(\boldsymbol{z}|\boldsymbol{z}')\,c(\boldsymbol{z}),\\
\big\{L(\boldsymbol{z}),\bar{c}(\boldsymbol{z}')\big\}&=+\I\,\delta^{(2)}(\boldsymbol{z}|\boldsymbol{z}')\,\bar{c}(\boldsymbol{z}),
\end{align}\end{subequations}
with $\boldsymbol{z}=(\vartheta,\varphi)$ and
\begin{subequations}\begin{align}
\big\{c(\boldsymbol{z}),U(\boldsymbol{z}')\big\}&=XU(\boldsymbol{z})\,\delta^{(2)}(\boldsymbol{z}|\boldsymbol{z}'),\\
\big\{\bar{c}(\boldsymbol{z}),U(\boldsymbol{z}')\big\}&=\bar{X}U(\boldsymbol{z})\,\delta^{(2)}(\boldsymbol{z}|\boldsymbol{z}'),\\
\big\{L(\boldsymbol{z}),U(\boldsymbol{z}')\big\}&=-\frac{1}{2}JU(\boldsymbol{z})\,\delta^{(2)}(\boldsymbol{z}|\boldsymbol{z}').
\end{align}\end{subequations}

The phase space is slightly larger than the physical phase space obtained from imposing the constraints \eref{ham-cons} and \eref{hol-eq}. There is one residual pair of second-class constraints,
\begin{align}
c\bar{a}\bar{b}&=-\sqrt{2}\gamma\left(L+\bar{a}a\right)\,\sqrt{\bar{a}a\bar{b}b}\,\tan\bigl(\sqrt{2\sigma\bar{\sigma}}\bigr)-\I \bar{a}a\bar{b}b,\label{scndclss}\\
\bar{c}{a}{b}&=-\sqrt{2}\gamma\left(L+\bar{a}a\right)\,\sqrt{\bar{a}a\bar{b}b}\,\tan\bigl(\sqrt{2\sigma\bar{\sigma}}\bigr)+\I \bar{a}a\bar{b}b.
\end{align}
The constraints are second-class. At the quantum level, only one of them can be imposed strongly, which leads to recurrence relations for the components of physical states. The other constraint maps the physical Hilbert space into its orthogonal complement. In the following, the $SL(2,\R)$ configuration variable $U$ plays no role in our analysis of the radiated power. It is enough to consider the subalgebra generated by $c$, $\bar{c}$, $L$, $a$, $\bar{a}$ and $b$, $\bar{b}$ alone.

\section{Critical luminosity}
\noindent Upon quantizing the oscillators $a,\bar{a}$,  and $b,\bar{b}$ and the $SL(2,\R)\times\mathfrak{sl}(2,\R)$ variables $(U,L,c,\bar{c})$, we obtain a kinematical Hilbert space. Physical states lie in the kernel of the constraint \eref{scndclss}. The kinematical Hilbert space carries a unitary representation of $SL(2,\R)$. Since the $SL(2,\R)$ Casimir commutes with the constraint, it will characterize physical states. If we express the Casimir in terms of geometric variables, we obtain
\begin{align}\nonumber
L^2-c\bar{c}&=\frac{1}{8\pi G}\biggl[\frac{1}{4\gamma^2}(E_--E_+)^2-
\frac{1}{\gamma^2}\sh^2\bigl(2\sqrt{\sigma\bar{\sigma}}\bigr)E_+E_-\\
&\quad-\frac{1}{2}(E_++E_-)^2\tan^2\bigl(\sqrt{2\sigma\bar{\sigma}}\bigr)\biggr].\label{Casmirdef}
\end{align}
When $L^2>c\bar{c}$, the spectrum of the Casimir is discrete. When $L^2<c\bar{c}$, it is continuous. For $\sigma=0$ and $E_+\neq E_-$, the Casimir is positive, i.e.\ $L^2>c\bar{c}$. As we increase $\sigma$, we will reach a critical value $\sigma_{\mtext{crit}}$, where the sign will flip. Expanding the Casimir  \eref{Casmirdef} for small shear, we obtain
\begin{equation}
|\sigma_{\mtext{crit}.}|^2=\frac{1}{4}\frac{(E_--E_+)^2}{\gamma^2(E_++E_-)^2+4E_+E_-}+\mathcal{O}(|\sigma_{\mtext{crit}}|^3).\label{critshear1}
\end{equation}
This equation determines the critical shear on a finite null surface in terms of the area densities $E_\pm$ at the two cuts at which the pulse starts and terminates. To connect this local bound with standard observables at infinity, we consider the asymptotic expansion with respect to an affine radial Bondi coordinate $r$ \cite{Bondi21,Sachs103,BONDI:1960aa} in a neighbourhood of future null infinity $\mathcal{I}^+$.  The shear of the ingoing null generators vanishes as $\mathcal{O}(r^{-1})$. The area density $\Omega^2$ blows up as $\mathcal{O}(r^2)$. Hence, $E_\pm(\vartheta,\varphi)\equiv\Omega^2(\mathcal{U}=\pm 1,\vartheta,\varphi)=\mathcal{O}(r^2)$.\smallskip

\begin{figure}[h]
\begin{center}
\hspace{1.5em}\includegraphics[width=0.35\textwidth]{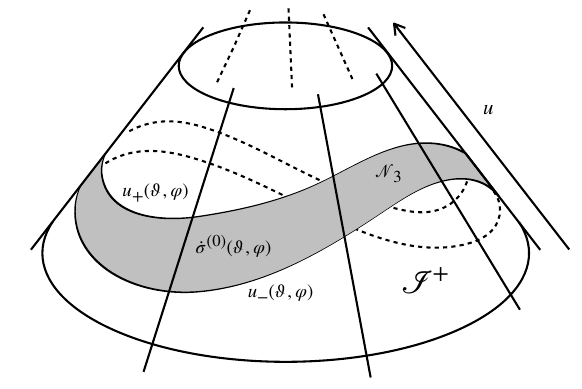}
\end{center}
\caption*{Figure 2: Impulsive data at future null infinity. In the shaded region $\mathcal{N}_3$, the time derivative of the asymptotic shear ${\sigma}^{(0)}(u,\vartheta,\varphi)$ is constant in $u$, everywhere else $\dot{\sigma}^{(0)}=0$. }\label{fig2}
\end{figure}

 To obtain the critical shear $\sigma_{\mtext{crit}}$ for an asymptotic rather than local observer, we proceed in three steps. First, we consider a pulse of gravitational radiation in terms of standard Bondi coordinates at future null infinity. Then, we translate the parameters that characterize this asymptotic pulse back into our local variables. The third step is to combine the results and compute the critical shear \eref{critshear1} for an asymptotic observer. Below, we have a pictorial representation of the setup at infinity (\hyperref[fig2]{Figure 2}). 

Consider thus a pulse of radiation at future null infinity $\mathcal{I}^+$ in terms of standard Bondi coordinates $(u,r,\vartheta,\phi)$. The pulse starts at an asymptotic Bondi time $u_-(\vartheta,\varphi)$ and terminates at $u_+(\vartheta,\varphi)$ with total duration $(\Delta u)(\vartheta,\varphi)=u_+(\vartheta,\varphi)-u_-(\vartheta,\varphi)$. During the duration of the pulse, the time derivative of the asymptotic Bondi shear ${\sigma}^{(0)}$ is constant, i.e.\ $\dot{\sigma}^{(0)}\equiv\dot{\sigma}^{(0)}(\vartheta,\varphi)$.\smallskip

 To leading order in the $1/r$-expansion,  the map between the boundary intrinsic time coordinate $\mathcal{U}:\mathcal{N}_3\rightarrow[-1,1]$ and the asymptotic Bondi time  $u$ is a mere angle-dependent dilation,
\begin{equation}
\partial^a_{\mathcal{U}}=\frac{(\Delta u)(\vartheta,\varphi)}{2}\partial^a_{u}+\mathcal{O}(r^{-1}).\label{Uueq}
\end{equation}
Upon introducing an adapted Newman--Penrose tetrad $(k^a,\ell^a,m^a,\bar{m}^a)$, see \cite{newmanpenrose}, where $k^a=\partial^a_r$ and $\ell^a=\partial^a_u+\mathcal{O}(r^{-1})$, and both $k_a=-\nabla_au$ and $\ell_a$ are surface orthogonal. Using the asymptotic falloff conditions, we obtain for the shear of the two null congruences that
\begin{align}
\sigma_{(\ell)}&:=m^am^b\nabla_a\ell_b=-\frac{\dot{\sigma}^{(0)}(u,\vartheta,\varphi)}{r}+\mathcal{O}(r^{-2}),\label{ell-shear}\\
\sigma_{(k)}&:={m}^a{m}^b\nabla_ak_b=\frac{\sigma^{(0)}(u,\vartheta,\varphi)}{r^2}+\mathcal{O}(r^{-3}).
\end{align}
To connect the Bondi parametrisation of the impulsive data with our framework, we identify $\mathcal{N}_3$ with a portion of $\mathcal{I}^+$.  The map \eref{Uueq} between the two clock variables determines the map between the asymptotic Bondi shear $\dot{\sigma}^{(0)}$ and the shear $\sigma$ at the abstract boundary $\mathcal{N}_3$ as defined in \eref{hol-eq}. We obtain
\begin{equation}
\sigma(r;\vartheta,\varphi)=\frac{(\Delta u)(\vartheta,\varphi)}{2}\frac{\dot{\sigma}^{(0)}(\vartheta,\varphi)}{r}+\mathcal{O}(r^{-2}).\label{sigmaUB}
\end{equation}
In this way, we generate an entire $r$-parameter family of radiative data $\{\sigma(r)\}_{r}$ on the abstract null boundary $\mathcal{N}_3$. As we go through different values of $r$, we generate different spacetime geometries in the bulk. As we send $r\rightarrow\infty$, the abstract boundary $\mathcal{N}_3$ turns into an asymptotic boundary. \smallskip

Next, we return to the critical shear \eref{critshear1} and translate $\sigma_{\mtext{crit.}}$ into a critical value for the asymptotic  value of $\dot{\sigma}^{(0)}$. This requires a further ingredient, namely the asymptotic Bondi expansion of the area element $\Omega^2$. Using the standard round metric at future null infinity $\mathcal{I}^+$, we have $\Omega^2(u,r,\vartheta,\varphi)=r^2-2(u-u_o)r+\mathcal{O}(r^0)$. This expansion implies for $\Omega^2\big|_{\mathcal{U}=\pm 1}=E_\pm$ that
\begin{align}
&E_+(\vartheta,\varphi)=E_-(\vartheta,\varphi)+\mathcal{O}(r^1)=r^2+\mathcal{O}(r^1),\label{E-expans1}\\
&E_-(\vartheta,\varphi)-E_+(\vartheta,\varphi)=2r(\Delta u)(\vartheta,\varphi)+\mathcal{O}(r^0).\label{E-expans2}
\end{align}
We insert the expansion back into \eref{critshear1} and obtain the critical shear of the null generators $\partial_{\mathcal{U}}^a$
\begin{equation}
|\sigma_{\mtext{crit.}}(r;\vartheta,\varphi)|^2=\frac{1}{4}\frac{(\Delta u)^2(\vartheta,\varphi)}{\gamma^2+1}\frac{1}{r^2}+\mathcal{O}(r^{-3}).\label{critshear2}
\end{equation}
Notice that the critical shear has the same characteristic $1/r$ falloff as $\sigma_{(\ell)}$, see \eref{ell-shear}. This is an immediate consequence of \eref{critshear1}, where we saw that the critical shear at finite distance has an implicit dependence on $E_\pm$, whose asymptotic expansion is given in \eref{E-expans1} and \eref{E-expans2}. We are now left to translate \eref{critshear2} back into the asymptotic Bondi frame using equation \eref{sigmaUB}. The resulting critical shear $\sigma_{\mtext{crit.}}$ corresponds to a critical value for the time derivative of the asymptotic Bondi shear given by
\begin{equation}
|\dot{\sigma}^{(0)}_{\mtext{crit.}}|=\frac{1}{\sqrt{\gamma^2+1}}+\mathcal{O}(r^{-1}).
\end{equation}
Using the Bondi mass loss formula, we infer a critical value for the luminosity of the gravitational wave pulse,
\begin{equation}
\mathcal{L}_{\mtext{crit.}}=\frac{c^5}{4\pi G}\oint_{S_2}d^2\Omega\,|\dot{\sigma}^{(0)}_{\mtext{crit.}}|^2=\frac{\mathcal{L}_{\mathrm{P}}}{\gamma^2+1}.\label{Lcrit}
\end{equation}


If $L^2-c\bar{c}> 0$, the impulsive wave will have a luminosity smaller than $\mathcal{L}_{\mtext{crit.}}$ In this regime, the Casmir has a discrete spectrum \cite{Bargmann:1947}. Each light ray carries a discrete unitary representation of $SL(2,\R)$. The fundamental operators are
\begin{align}
&c^\dagger|N,n_a,n_b,n_c\rangle=\I\sqrt{(N+n_c)(n_c-N+1)}|N,n_a,n_b,n_c+1\rangle,\\
&\left(L^2-\tfrac{1}{2}(cc^\dagger+c^\dagger c)\right)|N,n_a,n_b,n_c\rangle=N(N-1)|N,n_a,n_b,n_c\rangle,\\
&a^\dagger|N,n_a,n_b,n_c\rangle=\sqrt{n_a+1}|N,n_a+1,n_b,n_c\rangle,\\
&b^\dagger|N,n_a,n_b,n_c\rangle=\sqrt{n_b+1}|N,n_a,n_b+1,n_c\rangle,
\end{align}
where $n_a$ and $n_b$ are integers, $n_c=N,N+1,\dots$ is integer or half-integer and $N=1,\tfrac{3}{2},\dots$. Physical states are annihilated by the constraint \eref{scndclss}. For the discrete series representations of $SL(2,\R)$, a unique solution can be found by a linear combination of states
\begin{align}
\bigl\{|N,n+r,r,N+m-r\rangle:r=0,\dots,m\bigr\}.
\end{align}
If we held the Casimir fixed, the quantum number $m$ determines the difference $E_--E_+$, which is a Dirac observable, see \cite{Wieland:2024dop}.\smallskip 

For the continuous series representations of $SL(2,\R)$, $L^2-c\bar{c}<0$. The spectrum of the Casimir is continuous \cite{Bargmann:1947}, for any $s\in\R$, we have
\begin{align}
&c^\dagger|s,n_a,n_b,n_c\rangle=\frac{1}{2\I}(\I s-2n_c-1)|s,n_a,n_b,n_c+1\rangle,\\
&\left(L^2-\tfrac{1}{2}(cc^\dagger+c^\dagger c)\right)|s,n_a,n_b,n_c\rangle=-\frac{1}{4}(s^2+1)|s,n_a,n_b,n_c\rangle.
\end{align}
 In this regime, the impulsive wave will have a luminosity bigger than $\mathcal{L}_{\mtext{crit.}}$ The spectrum of the Casimir is continuous, and the operator $L$ is no longer bounded from below. This has important consequences. To illustrate the resulting state space, consider the recurrence relations under the simplifying assumption that $\gamma\rightarrow 0$.\footnote{The physical significance of this limit is as follows. Up to a boundary term, we can write the $\gamma$-term in the action as the square $T_\alpha\wedge T^\alpha$ of the torsion two-form $T^\alpha=\nabla e^\alpha$. The Barbero--Immirzi parameter, which appears in the denominator in front of $T_\alpha\wedge T^\alpha$, controls the width of certain fluctuations of $T^\alpha$. As $\gamma\rightarrow 0$, these quantum fluctuations of the torsion two-form are suppressed, see \cite{Freidel:2005sn,Daum:2013fu}. }  To solve the constraint, we start from a superposition of kinematical states
\begin{equation}
 \Psi_{s,n,m}=\sum_{r=0}^\infty \psi_m \big|s,n+r,r,m-r\big\rangle.\label{Psiphysdef}
\end{equation}
This state satisfies $ca^\dagger b^\dagger\Psi=-\I a^\dagger a b^\dagger b\Psi$ provided
\begin{equation}
\psi_{r+1}=\frac{\I}{2}\frac{s+\I\left(2(r-m)+1)\right)}{\sqrt{n+r+1}\sqrt{r+1}}\psi_r.
\end{equation}
The state $\Psi$ is normalizable as long as $n+2m>0$, in which case $\sum_{r=0}^\infty|\psi_r|^2={}_2F_1(\frac{1}{2}-m-\frac{\I}{2}s,\frac{1}{2}-m+\frac{\I}{2}s;n+1;1)\psi_o$, with ${}_2F_1(a,b;c;z)$ denoting the standard hypergeometric functions. 
 The main difference to the discrete series representations is that the recurrence relations no longer terminate.  Let us discuss possible consequences of this observation.\smallskip
 
In quantum gravity, there is no preferred time and no preferred Hamiltonian \cite{Rovelli:2015gwa}. Instead of the $S$-matrix, we expect to have \emph{local amplitudes} in finite regions  of spacetime  \cite{Oeckl2003,alexreview,Reisenberger:2000zc,lqcspinfoam2,Hartle:1983ai}. Local amplitudes are linear functionals $A[\psi]$ on the kinematical boundary Hilbert space, whose elements $\psi$ characterize kinematical data at the closed boundary of the underlying region. The usual probability interpretation extends to this formalism and provides frequencies for the repeated occurrence of the same physical process within a family of equally prepared experiments \cite{Oeckl:2006rs}. The local amplitudes can be built formally from the projector $\boldsymbol{P}$ of kinematical boundary states onto the physical Hilbert space through $A[\psi]=\mathrm{Tr}(\boldsymbol{P}\psi)$, see \cite{Oeckl2003,alexreview,Reisenberger:2000zc,lqcspinfoam2}. 
 We have seen in above, how one can, in principle, construct physical states on a finite null boundary. Thus, we implicitly have a candidate at hand for how to build the projector onto physical states. This in turn defines a proposal for the amplitude map, which contains contributions from both the discrete series and continuous series representations of $SL(2,\R)$. In our opinion, there is, however, an indication to remove the continuous series representations from the definition of the projector. The the physical states \eref{Psiphysdef} are superpositions of kinematical states, where the quantum numbers $n_a$ and $n_b$ will become arbitrarily large. For any such state, the shear $\sigma\bar{\sigma}$ will be unbounded from above, see \eref{shear-def}. This creates the following problem. If we include the continuous series representations and compute the probabilities for the shear $\sigma\bar{\sigma}$ taking a certain value, there will always be a chance that there are transitions in which one obtains a final state in which $\sqrt{2\sigma\bar{\sigma}}>\pi/2$. In this case, the profile of the area density \eref{Omprofile} passes through a caustic, where $\Omega^2=0$. When there is a caustic, we violate the implicit assumption that we are in a smooth asymptotic region, in which $\Omega^2=r^2+\mathcal{O}(r)$, as $r\rightarrow\infty$. This can be avoided only when we exclude the states from the amplitude map for which $L^2<c\bar{c}$. In this case, the luminosity of a gravitational wave pulse will always be bounded by $\mathcal{L}_{\mtext{crit.}}$. 
 A detailed analysis of the continuum limit is beyond the scope of this article.  
More work is required to decide whether the super-Planckian states that carry a luminosity greater than $c^5/G$ should be excluded from the physical Hilbert space.

 \section{Outlook}

\noindent  The critical luminosity \eref{Lcrit} separates the continuous spectrum from the discrete eigenvalues of the $SL(2,\R)$ Casimir. The bound depends on the Barbero--Immirzi parameter $\gamma$. This dependence is a common feature in $D=4$. 
When $\gamma\neq0$, the asymptotic boundary charges are a mixture of electric and magnetic contributions that otherwise vanish in the $\gamma\rightarrow\infty$ limit \cite{Wieland:2017cmf,Wieland:2021vef,Wieland:2024dop}. The spectrum of the charges is determined by their algebraic properties alone, but the map between the charges and the physical observables depends on $\gamma$. In this way, the spectrum of physical observables can depend on $\gamma$. This is analogous to how the $\theta$-angle in quantum electrodynamics enters the Dirac quantization condition between magnetic and electric charges \cite{Witten:1979ey}. 
As explained in e.g.\ \cite{Wieland:2017cmf,Freidel:2020ayo}, this effect is responsible for the discrete spectra of geometric observables  in loop quantum gravity \cite{Rovelliarea,AshtekarLewandowskiArea,ashvolume,bianchilength,bianchisommer}. Such a fundamental quantum discreteness of geometry may affect other physical observables. It can create a fundamental bound on the energy density of matter \cite{Ashtekar:2006rx,Ashtekar:2006wn} and perhaps also acceleration \cite{Rovelli:2013osa}. 
Here, we found  indications for the existence of a  similar bound on the gravitational wave luminosity \eref{Lcrit}. To the best of our knowledge no such mechanism has been explored in quantum gravity before. The analysis is based on a non-perturbative quantization of radiative data at finite distance. The result is only partial, because there is an implicit assumption: the validity of the classical asymptotic $1/r$-expansion when applied to the spectrum of gravitational observables at finite null boundaries. If the luminosity exceeds the critical luminosity \eref{Lcrit}, this assumption may no longer be valid due to the possible creation of caustics. What we have shown so far is only a first step. Our analysis was based on a specific class of piecewise constant null-initial data. A more refined investigation will follow to understand the continuum limit of our model and the significance of the Planck power for the fine structure of the spectrum of the gravitational wave luminosity.\smallskip

 One of the most pressing problems ahead is to connect the spectrum of the radiated power to the quantum state of the radiating sources. Our entire approach is based on a quasi-local quantisation of the radiative data at the boundary of a finite domain. This makes the problem conceptually  different from an asymptotic quantisation, in which we keep the radial coordinate classical and send it to infinity before considering quantum effects. Our model, which disagrees with the standard asymptotic quantisation, in which the spectrum of the radiated power is continuous and unbounded from above, is, in fact, an implicit proof that the $r\rightarrow\infty$ limit may not commute with the Dirac quantization procedure. Consider the following scenario. Take the present model as a candidate theory for the quantisation of the radiative data on an abstract, but finite, null boundary. Choose within this theory a suitable family of coherent states $\{\Psi_r\}_{r\in\R_>}$ for classical radiative data on a family of null boundaries $\{\mathcal{N}_{r}\}_{r\in\R_>}$ embedded into a classical spacetime.\footnote{It is worth mentioning that for the discrete series representations of $SL(2,\R)$ such coherent states can be  constructed easily using the oscillator representation of the $\mathfrak{sl}(2,\R)$ algebra. }  In this way, we can generate a foliation in which the $r\rightarrow\infty$ limit sends us to future (past) null infinity. Our results show that by taking the $\hbar\rightarrow0$ limit first, then sending $r\rightarrow\infty$, we may not return to the standard radiative phase space  \cite{AshtekarNullInfinity} in which the luminosity is unbounded from above. If this scenario is true, whatever semi-classical states we choose, we would always land in a portion of phase space, where the luminosity is bounded. Moving forward, we may find it necessary to extend the usual $S$-matrix theory into a framework of quasi-local amplitudes. Such a reasoning resonates with the idea of using boundary conditions for incoming radiation at so-called \emph{finite infinity} \cite{Ellis1984,Wiltshire:2007jk} to describe isolated systems and trace back how the large scale structure can affect local measurements of a freely falling observer in a cosmological spacetime. It will be important to understand whether the type of $\hbar\neq 0$ effect studied in this paper, can then also alter the spectrum of other classical observables. For example, it seems that our techniques can be applied to the Einstein–Yang–Mills system as well. It will be interesting to investigate the combined effect of the $\theta$-angle and the Barbero–Immirzi parameter $\gamma$ on the non-perturbative quantization of the coupled system.\smallskip

\noindent{\emph{Acknowledgments.}  This research was funded in part through a Heisenberg fellowship of Deutsche Forschungsgemeinschaft (DFG, German
Research Foundation)---543301681.}

\providecommand{\href}[2]{#2}\begingroup\raggedright\endgroup

\end{document}